\documentclass[manuscript]{aastex62}

\usepackage{CJKutf8}
\usepackage[english]{babel}
\usepackage[T1]{fontenc}
\usepackage{amsmath}    %equation case
\usepackage{booktabs}
\usepackage{multirow}
\usepackage{footmisc}
\revised{\today}
\submitjournal{\pasp}

\shortauthors{Lee et al.}

\begin{document}

%%\tracingall
\begin{CJK}{UTF8}{mj}

%\title{Center Finding Algorithm for Point Source Observation in Slit Camera Package of IGRINS}
%\title{CENTER FINDING ALGORITHM OF SLIT-VIEWING CAMERA SOFTWARE}
\title{IGRINS Slit-Viewing Camera Software}

\email{hyeinlee@khu.ac.kr} 

\author{Hye-In Lee}
\affil{School of Space Research, Kyung Hee University, 1732 Deogyeong-daero, Giheung-gu, Yongin-si, Gyeonggi-do, 17104, Republic of Korea}

\author{Soojong Pak\correspondingauthor{}}
\affil{School of Space Research, Kyung Hee University, 1732 Deogyeong-daero, Giheung-gu, Yongin-si, Gyeonggi-do, 17104, Republic of Korea}

\author{Gregory N. Mace}
\affiliation{Department of Astronomy, University of Texas at Austin, Austin, TX 78712, USA}

\author{Kyle F. Kaplan}
\affiliation{SOFIA Science Center, NASA Ames Research Center, Moffett Field, CA 94035, USA}

\author{Huynh Anh N. Le}
\affiliation{Department of Astronomy, University of Science and Technology of China, Hefei, People's Republic of China}

\author{Heeyoung Oh}
\affiliation{Korea Astronomy and Space Science Institute, Daejeon 34055, Republic of Korea}

\author{Chan Park}
\affiliation{Korea Astronomy and Space Science Institute, Daejeon 34055, Republic of Korea}

\author{Sungho Lee}
\affiliation{Korea Astronomy and Space Science Institute, Daejeon 34055, Republic of Korea}

%%-------------------------------------------------------------------------------------------------------------
\begin{abstract}
We have developed observation control software for the Immersion GRating INfrared Spectrometer (IGRINS) slit-viewing camera module, 
which maintains the position of an astronomical target on the spectroscopic slit.
It is composed of several packages that monitor and control the system, acquire the images, and compensate for the tracking error by sending tracking feedback information to the telescope control system.
For efficient development and maintenance of each software package, we have applied software engineering methods, i.e., a spiral software development with model-based design.
It is not trivial to define the shape and center of astronomical object point spread functions (PSFs), which do not have symmetric Gaussian profiles in short exposure ($<$4 s) guiding images.
Efforts to determine the PSF centroid are additionally complicated by the core saturation of bright guide stars.
We have applied both a two-dimensional Gaussian fitting algorithm (2DGA) and center balancing algorithm (CBA) to identify an appropriate method for IGRINS in the near-infrared $K$-band.
The CBA derives the expected center position along the slit-width by referencing the spillover flux ratio of the PSF wings on both sides of the slit.
In this research, we have compared the accuracy and reliability of the CBA to the 2DGA by using data from IGRINS commissioning observations at McDonald Observatory.
We find that the performance of each algorithm depends on the brightness of the targets and the seeing conditions, with the CBA performing better in typical observing scenarios.
The algorithms and test results we present can be utilized with future spectroscopic slit observations in various observing conditions and for a variety of spectrograph designs.
\end{abstract}

%%-------------------------------------------------------------------------------------------------------------


%%-------------------------------------------------------------------------------------------------------------

\section{Introduction}
Slit or fiber spectrographs are the most commonly used in modern astronomical spectroscopy.
Since a slit or a fiber tip on the focal plane collects photons for dispersion, the signal to noise ratio (S/N) of a point target depends on the reliability of centering it in the collector.
%Additionally, observed data are influenced by thermal emission and atmospheric absorption of H$_{2}$O, CO$_{2}$, and etc. in the infrared (IR).
Studies by \citet{shimono12} and \citet{tamura12} showed the importance of well defined relationship between a system and the control software in the development of Prime Focus Spectrograph and the Fiber Multi Object Spectrograph for Subaru Telescope.
In the case of SpeX on the NASA Infrared Telescope Facility \citep{rayner03,rayner17}, the instrument control software consists of four components on Bigdog, Guidedog, and Littledog computers.
The Guidedog controls the infrared (IR) slit-viewing camera (SVC) to monitor the spillover photons from the target on the slit or the off-slit guide star in the field.
Since the refraction index in IR bands is smaller than that in visible bands, the IR image is less affected by the atmospheric turbulence and the IR guiding is more stable and efficient than in the optical \citep{lacy02,rayner03}.

We have developed the Slit Camera Package (SCP) for operation of the Immersion GRating INfrared Spectrometer (IGRINS) \citep{kwon12,chanpark14,mace16}.
IGRINS control software was designed with numerous center-finding algorithms in accordance with software engineering methods.
There were a number of complications that had to be overcome, like saturated centers and asymmetries of the point spread function (PSF) images in short exposures ($<$4 s) caused by atmospheric turbulence.
The asymmetric PSF profile can be a large fraction of a slit width.
Appropriate observation modes and scenarios can compensate for these properties and reduce slit-loss, which means more science target flux on the spectral detectors and higher S/N spectra.
We report the performance of a guiding algorithm that has not been used with previous IR spectrographs, so it can be useful for the future development of similar instruments.

In this paper, we describe IGRINS control software and the requirements of the SCP, including the center balancing algorithm (CBA). 
Section 2 explains the control software architecture design and the concepts of pointing and guiding for IR observations.
In Section 3, we show the results with the 16 sample guide star images from the IGRINS SVC. 
We then discuss the performance test results in Section 4, and discuss quantitative differences.
Finally, we will summarize all of this work in Section 5.

%%-------------------------------------------------------------------------------------------------------------

\section{Overview of the SCP}

IGRINS is a compact and high-resolution ($R$ = 45,000) near-IR, cross-dispersed, echelle spectrograph.
In 2014 March, IGRINS had the first commissioning observations on the 2.7~m Harlan J. Smith telescope (HJST) at McDonald Observatory \citep{chanpark14}.
The optical design of IGRINS was carefully optimized to have high throughput. 
With simultaneous exposures in $H$- and $K$-bands, IGRINS observes continuous wavelength spectra from 1.45 to 2.5 $\micron$ using a silicon immersion grating \citep{jaffe98,marsh07,wang10,gully12}, which was more than 30 times the spectral grasp of the original CRIRES on the Very Large Telescope \citep{kaufl04}.
The spectral S/N, per resolution element, of a point source observed with IGRINS in the $K$-band from the 2.7~m HJST can be greater than 100 for $K$ = 10.3 mag targets \citep{mace16}. 
IGRINS has also been commissioned on Lowell Observatory's 4.3~m Discovery Channel Telescope and the 8.1~m Gemini South Telescope \citep{mace18}.  

IGRINS employs two science grade Teledyne H2RG HgCdTe detectors for the $H$- and $K$-band spectrograph channels.
The SVC has an engineering grade H2RG placed behind a $K$-band filter, providing both target acquisition and guiding.
On the HJST, the SVC field of view (FoV) is 1$\arcmin$.9 $\times$ 3$\arcmin$.1 and the slit size is 1$\arcsec$.0 $\times$ 14$\arcsec$.8 with the plate scale of 0$\arcsec$.12 per pixel.
To achieve a high throughput, the slit width of IGRINS was designed to be approximate the same as the full width at half maximum (FWHM) of a point source PSF in a typical seeing conditions at the HJST \citep[$\sim$1$\arcsec$][]{chanpark14} 

\subsection{Architecture of IGRINS control software}
Typical software is developed via sequential progresses through software requirements analysis, design, code generation, testing, and commissioning processes.
This development method is called the linear sequential model, or waterfall model.
When the software requirements are not clearly defined at the beginning, e.g., for innovative scientific instruments, the development should follow the spiral model, in which the software evolves by iteration of the waterfall model \citep{pressman01}.
We applied the spiral model to our IGRINS software development.
The spiral model ensures that even after commissioning, the software can be adapted to the as-built performance and any new operation requirements.

\begin{figure}%%[ht!]
\epsscale{1.2}
\plotone{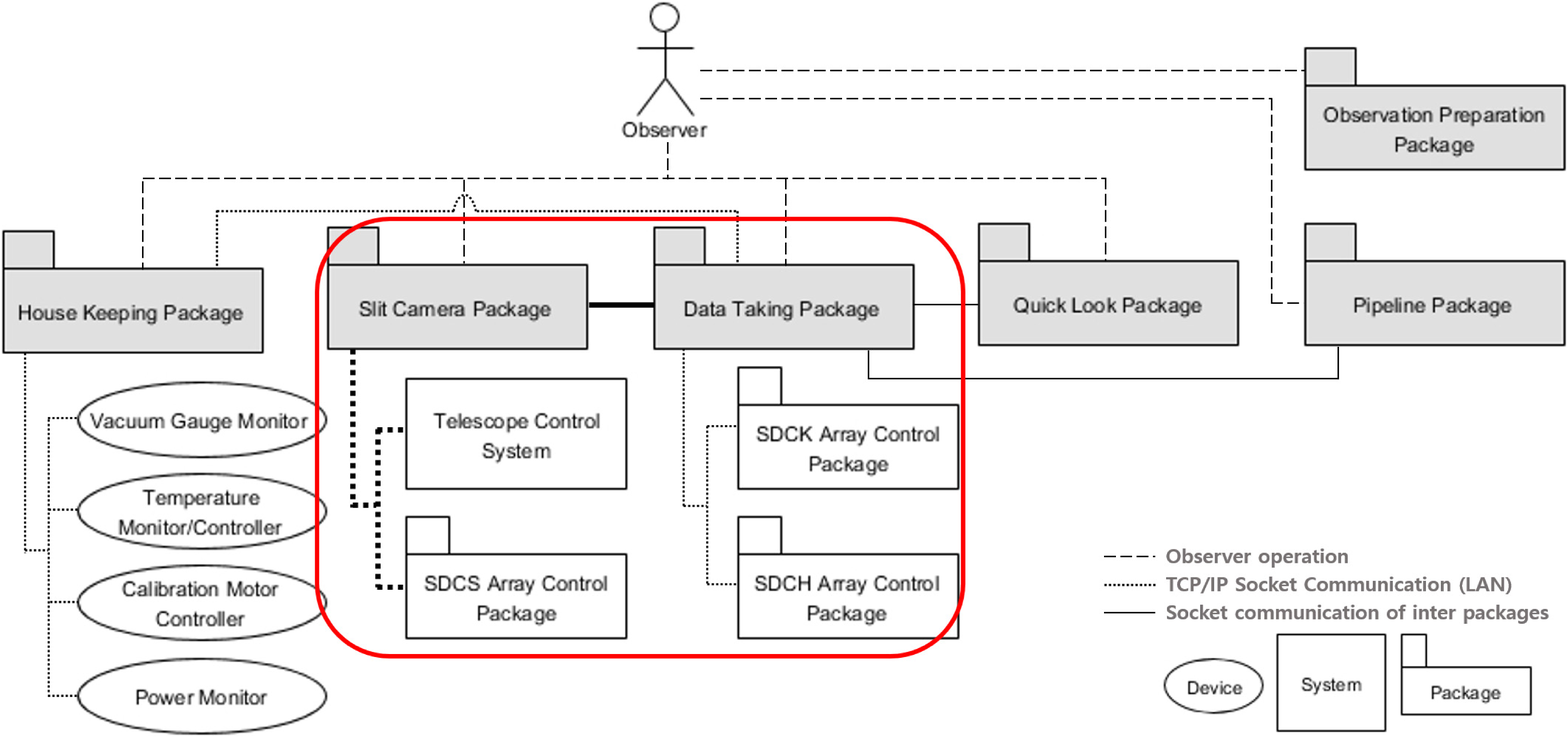}
\caption{IGRINS control software architecture and network connection. 
The pointing and guiding parts are enclosed within the red solid line. 
{The packages developed for IGRINS control software system are in the gray boxes: the House Keeping Package (HKP), the Slit Camera Package (SCP), the Data Taking Package (DTP), the Quick Look Package (QLP), the Pipeline Package (PLP), and Observation Preparation Package (OPP).
Other packages from the observatory or from the off-the-shelf are in the white boxes: the Telescope Control System (TCS), Science Detector Computer for Slit-camera (SDCS) Array Control Package, Science Detector Computer for $K$-band (SDCK) Array Control Package, and Science Detector Computer for $H$-band (SDCH) Array Control Package \citep{kwon12}.
We used Unified Modeling Language (UML) (Visual Paradigm 15.1, Visual Paradigm International Ltd.) to make this diagram.} \label{fig:network}}
\end{figure}

\begin{figure}
\epsscale{1.1}
\plotone{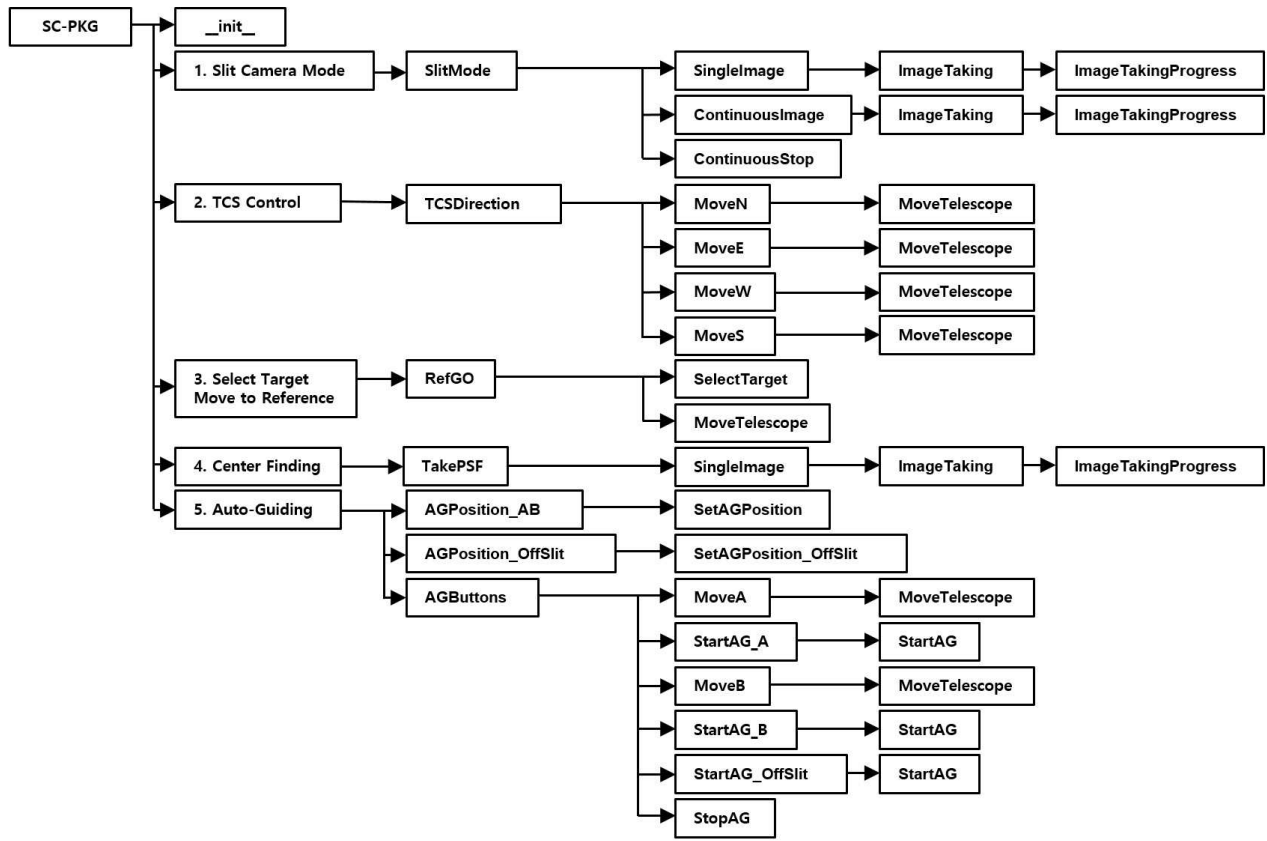}
\caption{Function mapping diagram for the Slit Camera Package (SCP). 
It shows functional properties following observation scenarios.
``SC-PKG'' means the Slit Camera Package (SCP), ``\underline{  }init\underline{  }'' means the initial function of (Graphic User Interface) GUI in Python code.
Numbers 1 to 5 are major functions defined by requirements. \label{fig:func_mapping}}
\end{figure}

\begin{figure}
\epsscale{1.2}
\plotone{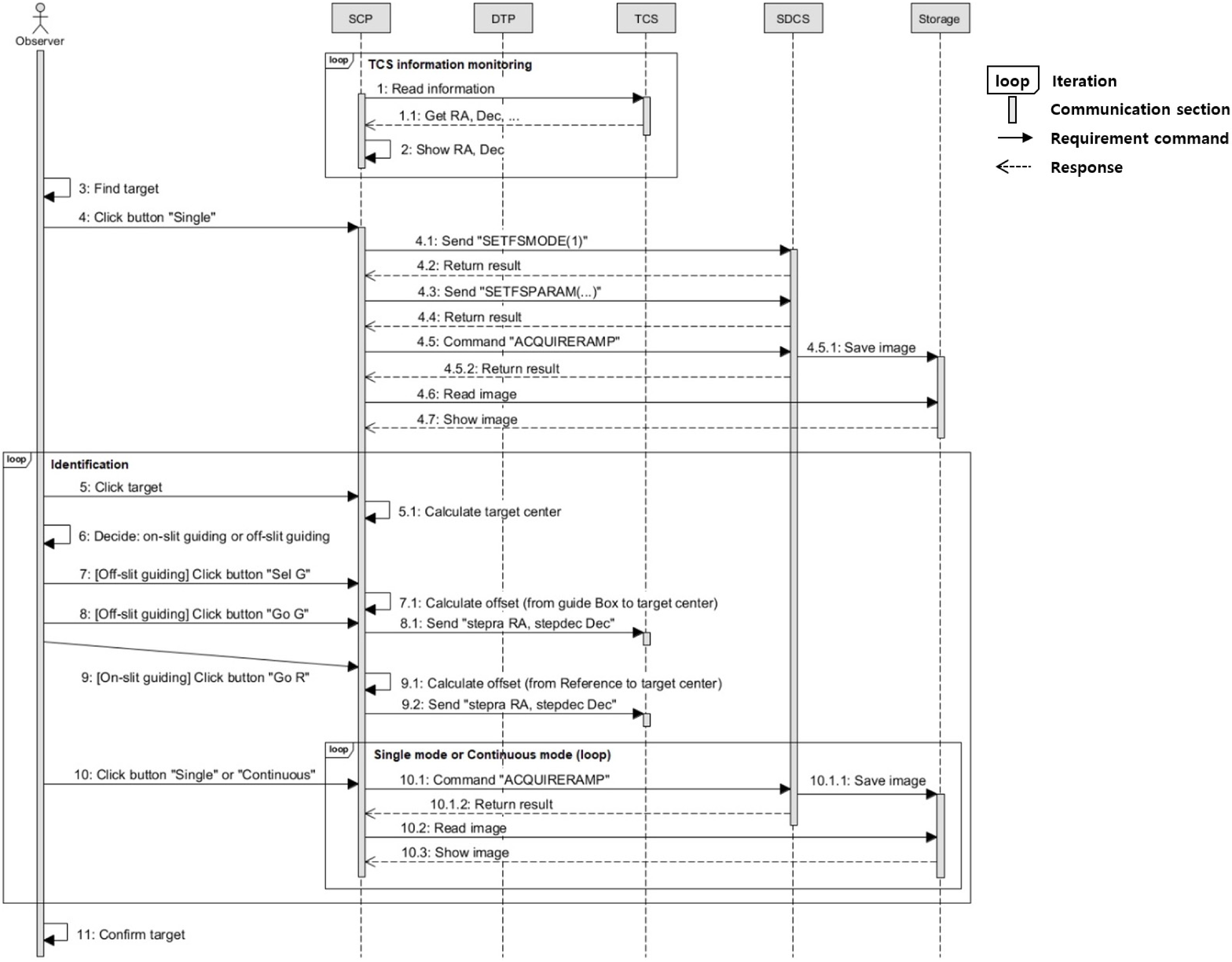}
\caption{Sequence diagram for target pointing. 
These processes show that ``3. Select Target Move to Reference'' of major functions in Figure \ref{fig:func_mapping}.
The sequential order numbers are labeled from 1 (read information) to 11 (confirm target):
1. The Slit Camera Package (SCP) requires information from the Telescope Control System (TCS). 
2. The SCP displays the TCS information (R.A. and decl.).
1 and 2 continue until finishing the current observation.
3. The observer finds a target by ``Single'' mode.
4. Then, the SCP communicates with SDCS for taking a data.
5. After finding the target, the SCP calculates the center of the target.
6-9. The observer decides and commands whether ``on-slit guiding'' or ``off-slit guiding'', the guiding star moves into a pre-defined box (A-box or guide box). 
10. The observer confirms to identify the target through ``Single'' or ``Continuous'' mode.
The auto-guiding process follows after this pointing process.
\label{fig:seq_TargetCentering}}
\end{figure}

The life time of astronomical instruments are so long, e.g., more than 10 yr, that the control software needs to be maintained and updated by software engineers who probably did not participate in the initial development.
In this case, the software should be modularized using object-oriented programming methods, and the structure of the functional elements should be clearly documented.
We designed IGRINS control software by using Unified Modeling Language (UML), which can visualize functional interactions in structure diagrams, e.g., architecture and network connection diagram (see Figure \ref{fig:network} for example) and function mapping diagram (see Figure \ref{fig:func_mapping} for example), and sequential operations in behavior diagrams, e.g., sequence diagram (see Figure \ref{fig:seq_TargetCentering} for example) \citep{balmelli07}.

IGRINS control software runs on the Instrument Control Computer, which is independent but networked to both IGRINS and the telescope control system (TCS).
It consists of the House Keeping Package (HKP), the Data Taking Package (DTP), and the SCP. 
Figure \ref{fig:network} shows IGRINS software architecture and the network connections \citep{rayner03, kwon12}.
Components in all packages were analyzed and designed by UML.
In addition to the sequential operations, they do threaded execution to avoid memory leak in the platform.
The HKP controls and monitors the status of the hardware components.
The DTP takes spectral data, provides approved observing modes, and controls the data flow and archiving.
It also interacts with the calibration module via the HKP.
The SCP controls the SVC and communicates with the TCS.
It performs an IR auto-guiding function either by the off-slit guide star in the slit camera field or by the spillover photons from the target on the slit \citep{rayner03,iseki08,landoni12}.
The software was written in Python 2.7 and used Tcl/Tk for graphic user interface (GUI) the elements on the environment of the Macintosh operating system.

\subsection{Operation Sequence}
In Figure \ref{fig:func_mapping}, we modeled every function and command using UML.
Each required element is to be a function code or command.
We arranged the major functions in the GUI. 
Each functions can have sub-functions, and some of the functions can be called by the major functions of high levels.
We deployed a sequence diagram for each operation (see example in Figure \ref{fig:seq_TargetCentering}).

\subsubsection{Boot-up and Shutdown}

After the SCP opens a socket for internal connection to the DTP, it loads the initial coefficients in the configuration file. 
The DTP and the SCP are connected with the science detector control computers for each H2RG array.
When the connections between the HKP and the DTP, the DTP and the SCP, and the SCP and the TCS are ready, the SCP sends command to the Science Detector Control Computer for Slit Camera (SDCS) to initialize the detector and to prepare for taking an image.
It also starts to get a current information of the TCS through TCP/IP.
When all observation processes are complete, all packages sequentially exit alive threads and the GUI in the reverse order of the boot-up processes.

\subsubsection{Pointing and Guiding}
\label{sec:pointing and guiding}

The concepts of pointing in the telescope control include identifying the science target and setting the target at the observing position, e.g., the spectroscope slit aperture \citep{mcgregor00}.
We defined a reference position (the green cross in Figure \ref{fig:scp}) which is near the slit in the SVC FoV \citep{chen12,mccormac13}.
After confirming the target on the reference position, the SCP finds the center of the target image and moves the telescope to place the target on the slit.

For faint or extended targets the center of the target is difficult to define and identify.
In these cases we use a guide star whose offset position from the target is accurately known.
Placing the guide star in the pre-set guide box defined from the offset (the pink box in Figure \ref{fig:scp}), we can efficiently and effectively confirm the target on the reference position and move the target on the slit.

While IGRINS acquires spectroscopic observations with exposure times between 1.63 and 1800 s, the SCP continuously takes the slit view images as \cite{rayner03} did for SpeX, and checks the pointing by finding the center of the target on the slit or the guide star in the guide box.
The telescope tracking errors can be compensated for by periodically commanding the TCS to adjust pointing.
The pointing process is shown in the sequence diagrams in Figure \ref{fig:seq_TargetCentering}.

\subsubsection{Observing Mode}
Infrared spectroscopy relies on periodically nodding the telescope pointing between the target and the blank sky \citep{lacy02,rayner03} to permit the subtraction of thermal background emission from the telluric atmosphere, the telescope mirrors, and the instrument optics.
To collect the background data in the same observing conditions, observers need to be able to adjust telescope pointing between spectroscopic frames.

\begin{figure}
\epsscale{0.8}
\plotone{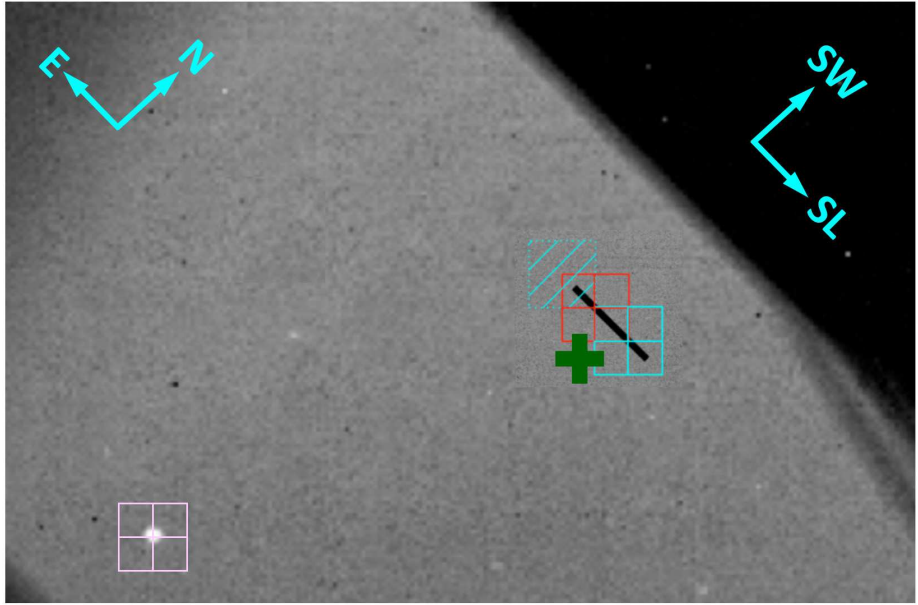}
\caption{Slit Viewer part of Graphic User Interface (GUI) of the Slit Camera Package (SCP). 
The green cross is the reference position, the red and cyan boxes are pre-defined A and B position of on the slit. 
The pink box is a guiding position for off-slit mode. 
The cyan dotted box filled of diagonal lines means blank sky position.
The user can turn on or off the guide box and blank sky box through checking ``Use Guide Position'' and ``Show Sky'' in the GUI, respectively.
\label{fig:scp}}
\end{figure}

We defined an A-box (the red box in Figure \ref{fig:scp}) for the target on the slit position and a B-box (the cyan box in Figure \ref{fig:scp}) for sky observations.
Both box positions are pre-determined by observers, and the pointing can be performed by centering the target inside the boxes.
When the target is at the A-box, we can expect to take the data from the target.
The reduction pipeline subtracts the B-frame, which was taken with the target at the center of the B-box, from the A-frame.
Note that the B-box position on the slit camera FoV does not correspond to the blank sky.
For observers to check the blank sky position, the SCP shows a ghost B-box (the cyan dotted box filled of diagonal lines in Figure \ref{fig:scp}) which is at the opposite side of the B-box.

For point source targets, observers can use a nod-on-slit mode, where both A-box and B-box are on the slit. 
Unless the target is too faint, the pointing calibration is done by taking a target image on the slit. 
In this mode, both A-frame subtracted by B-frame and B-frame subtracted by A-frame can be combined in the reduction process.

For angularly extended targets, a nod-off-slit mode, in which the A-box is on the slit and the B-box is out of the slit, should be used.
If the target has a well-defined peak at the center, e.g., active galactic nuclei, observers can use the target image on the slit to point.
Otherwise, observers should point the target by placing and centering the guide star at the guide box.
By using the user-defined script mode, with incremental offset values between the A-box and the guide-box, a slit-scan mode can produce a three-dimensional data cube.

\subsection{Center Finding Methods}
When the target is placed on the slit, most of the target photons pass through and are then collected by the spectrograph optics.
The slit camera image contains only the spillover photons, but the center of the PSF can be derived by fitting the spillover using the two-dimensional Gaussian fitting algorithm (2DGA; \citealp{press07}).
However, this maximum likelihood estimation can have fatal biased errors when the peaked profile is blocked by the slit.

\begin{figure}
\epsscale{.5}
\plotone{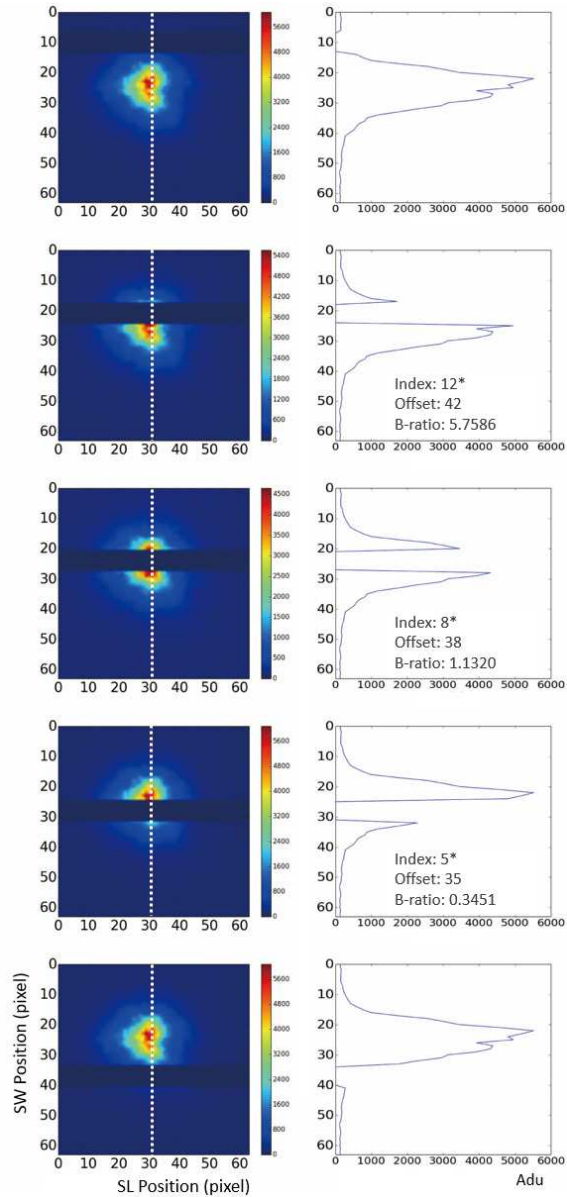}
\caption{Sample point source images overlapped by virtual slits.
The target is SS433 ($m_{K}$ = 8.2 mag) taken on 2014 May 24.
As Figure \ref{fig:D_LW}, it converts $X$ and $Y$ coordinate into slit-length and slit-width.
(``SW-position'' is slit-width direction, this is a relative position value (pixel) in pre-defined box on the reference position.)
The right plots show the one-dimensional profile along the slit width direction (marked with the white dashed line on the left image).
Inside the right plots, the ``Offset'' value shows the offset position of the virtual slit in pixel units.
``$B$-ratio'' is the spillover energy ratios of upper to lower parts from the virtual slits.
\label{fig:slitmask}}
\end{figure}

\begin{figure}
\epsscale{0.8}
\plotone{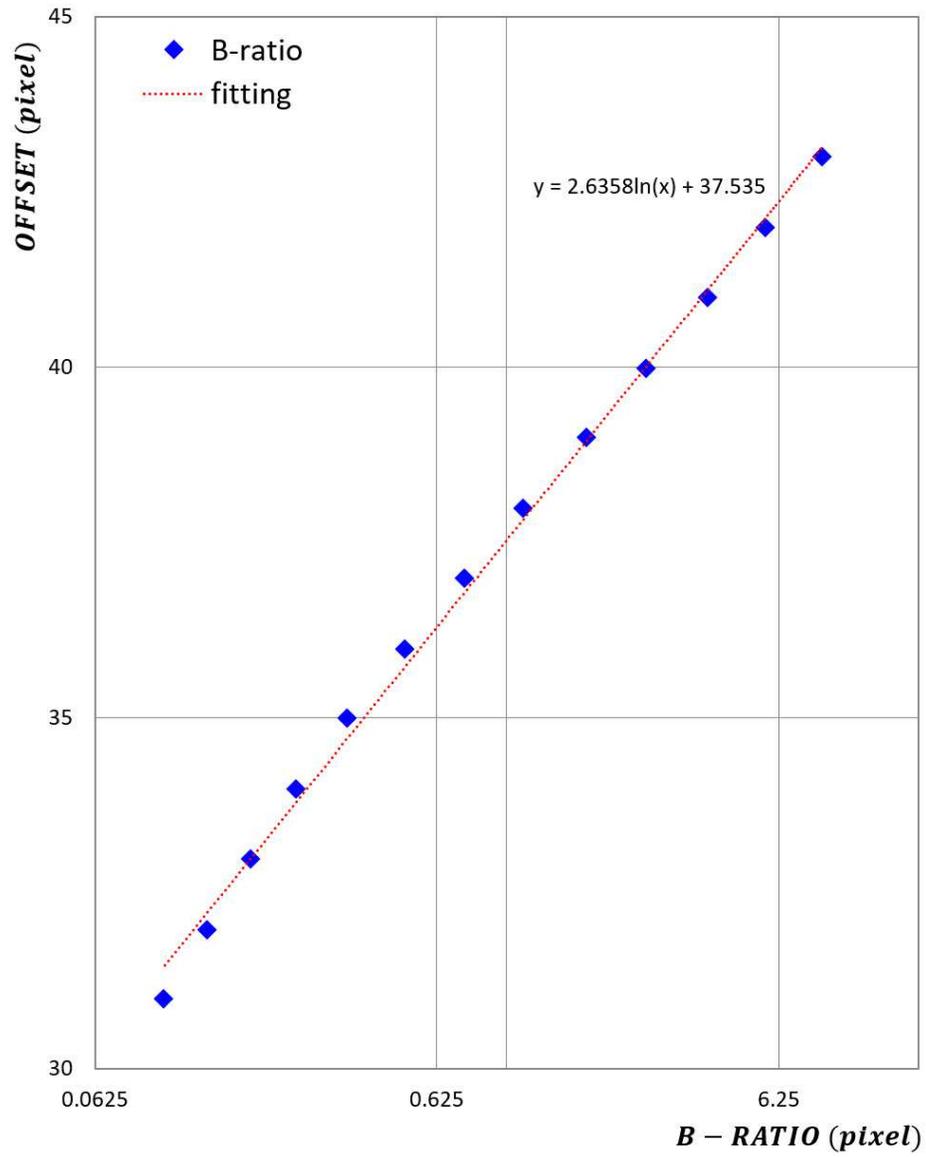}
\caption{Plot on a log scale of Offset vs. $B$-ratio.
The data are from Table \ref{tab:BT}.
We only consider the $B$-ratio value ranges which are larger than 0.1 and smaller than 10.
\label{fig:bt_logscale}}
\end{figure}

In the SCP, we developed a simple and robust CBA to define the center of the target.
Assuming the PSF is symmetric in the wings, the maximum throughput can be achieved when the spillover photons in the slit width direction are balanced \citep{rayner03,landoni12,rayner17}.
When observers identify and confirm the target on the reference position (see Section \ref{sec:pointing and guiding}), the target image without any slit obstruction can be obtained.
As shown in Figure \ref{fig:slitmask}, the SCP can put virtual slit blocks at various offsets and simulate the slit-obstructed PSF images with which we can derive the balance ratios ($B$-ratios) of the upper to the lower spillover flux values.
Table \ref{tab:BT} lists simulated $B$-ratios from the sample images in Figure \ref{fig:slitmask}.
Because the slit offset values and the $B$-ratios are mapped linearly in Figure  \ref{fig:bt_logscale}, the SCP can effectively estimate the offset position of the target from the measured $B$-ratio during observations.

\begin{figure}
\epsscale{1.2}
\plotone{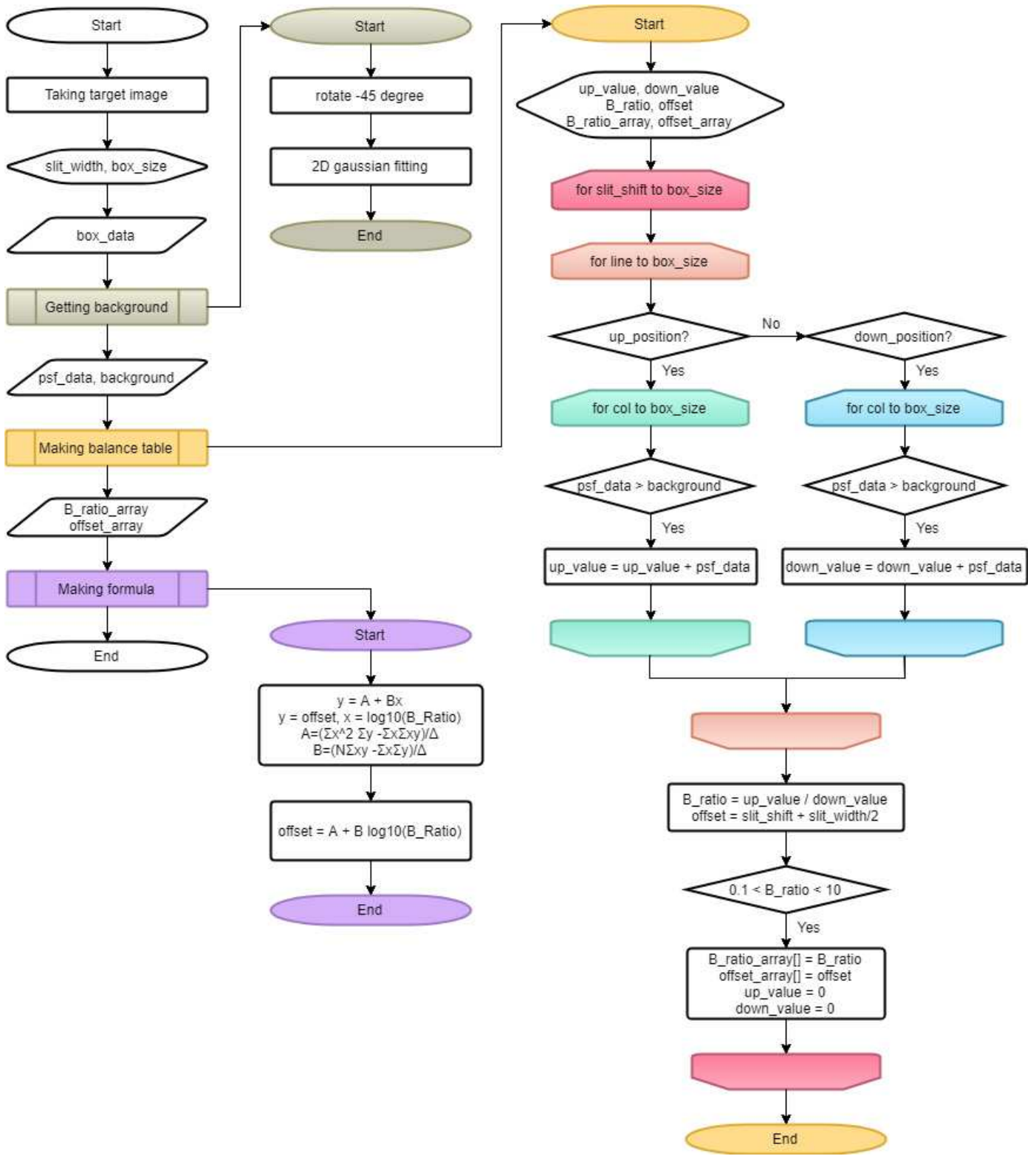}
\caption{Flowchart to make a balance table. 
The part of the pointing process at the reference position is included in the process of making balance table.
\label{fig:MakeBTFlowChart}}
\end{figure}

\begin{figure}
\epsscale{1.2}
\plotone{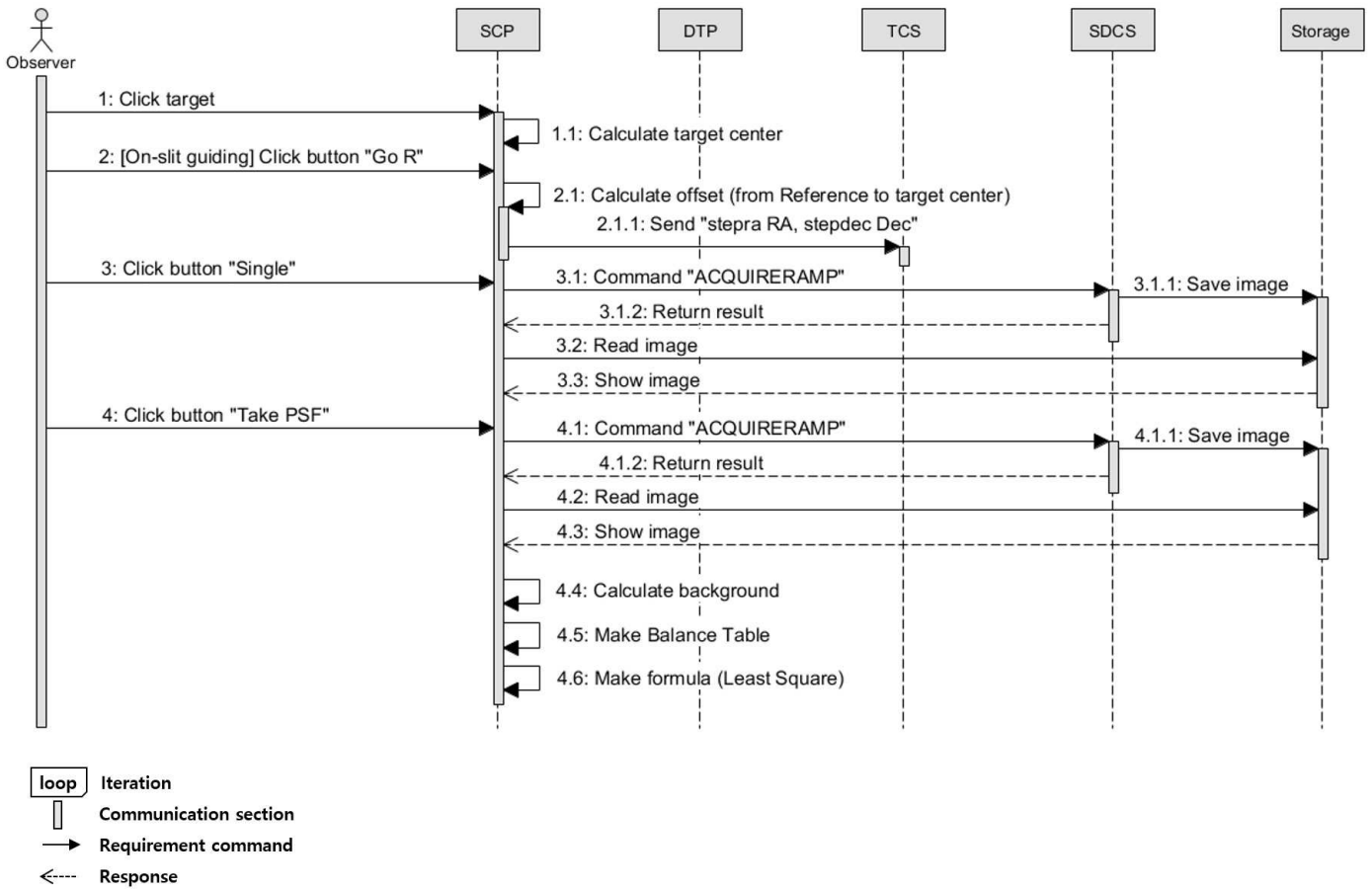}
\caption{Sequence diagram for a making balance table.
The sequential order numbers are labeled from 1 to 4:
1 and 2. The observer clicks the target and the button ``Go R''.
Then, the Slit Camera Package (SCP) calculates the center of the target and the offset from the current position to the reference position.
The SCP commands the Telescope Control System (TCS) to move the offset.
3. The observer confirms that the target is on the reference position by taking the image (``Single'' button)
4. and clicks ``Take PSF''.
The SCP makes a balance table and formula (offsets and ratios).
\label{fig:seq_make_BT}}
\end{figure}

In order to explain the CBA and the way of IGRINS software design, we show the flowchart (see Figure \ref{fig:MakeBTFlowChart}) and the sequence diagram (see Figure \ref{fig:seq_make_BT}) to make the balance table.
Using the flowchart we can analyze the requirements and derive the solution model.
Based on the flowchart, we included interactions with other software packages and designed the sequence diagram using UML.

\subsection{Auto-guiding Mode}
After the SCP creates the balance table from the target image on the reference position, observers can move the target into the A-box and begin acquiring spectra.
The SCP takes the point target image on the slit, and calculates the energy ratios ($B$-ratio) of the lower to the upper parts of the slit.
Applying the measured $B$-ratio value to the model relationship, the autoguider can move the target onto the center of the slit width by sending the position offset commands to the TCS.
In the auto-guiding mode, this process repeats, while the DTP takes spectroscopic data.

%%------------------------------------------------------------------------------------------------------------

\section{Performance Test}

\subsection{Sample Data}
The test data were taken at the McDonald 2.7~m telescope during IGRINS commissioning runs in 2014 March, May, and July.
We selected 618 SCP frames from 16 point sources (5 < $m_{K}$ < 9 mag) with various environmental conditions (see Table \ref{tab:result}).
In our sample frames, the positions of the point sources were distributed around the center of the slit width direction \citep{iseki08}.
Since the shortest integration time of the array readout electronics is 1.63 s \citep{ujjeong14}, bright targets in good seeing conditions were easily saturated at this short test integration time.

\subsection{Expected Target Position}
\begin{figure}
\epsscale{1.1}
\plotone{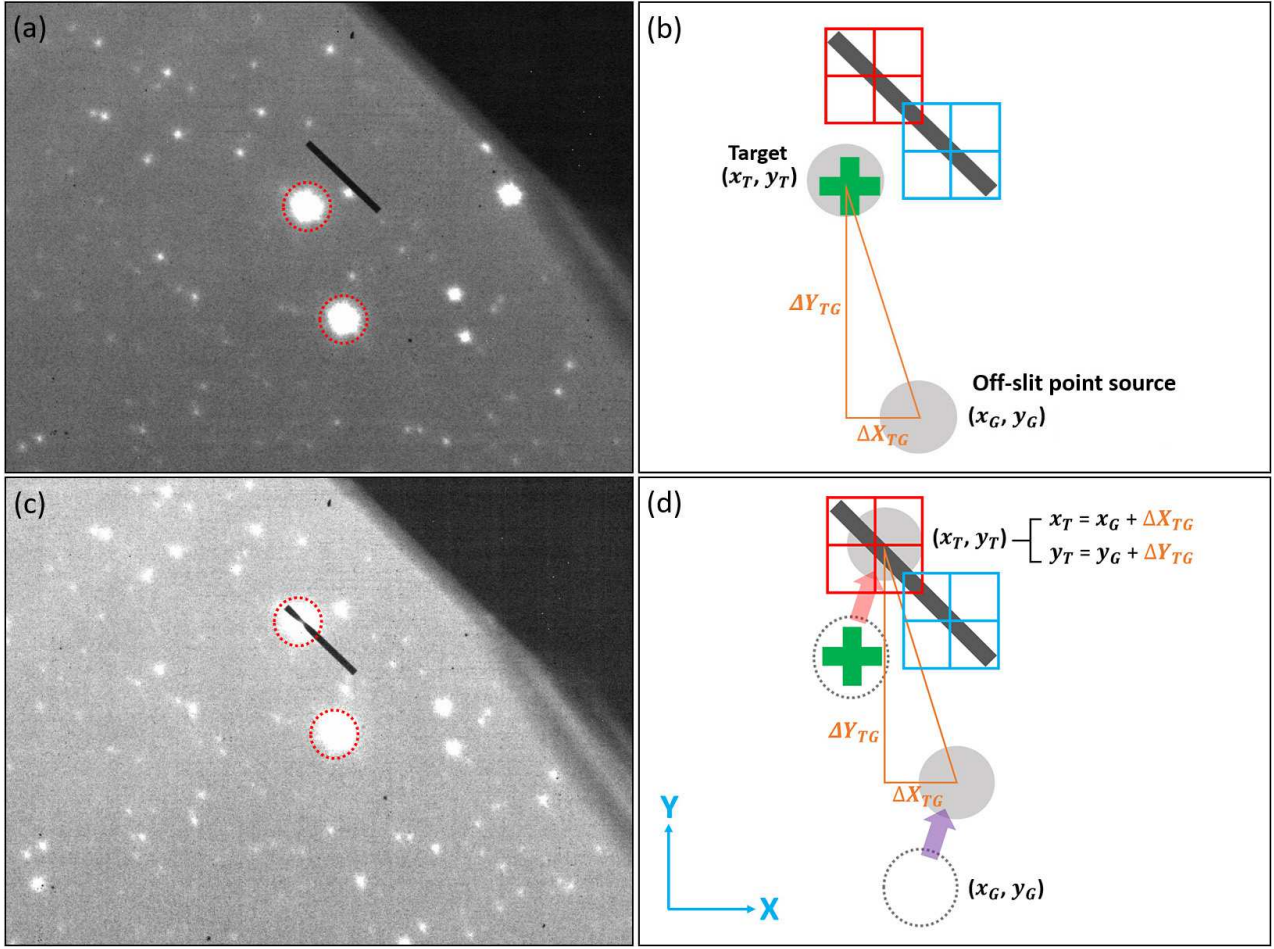} 
\caption{%Test process. 
Sample SCP images showing the expected target position.
The target is 2MASS J18331755 ($m_{K}$ = 6.8 mag) taken on 2014 May 25.
(a) and (b) show that the target is on the reference position (the green cross).
(c) and (d) show that the expected target position on the slit is derived from the off-slit point source (see the details in the text).
\label{fig:TestProcess}}
\end{figure}

When the target was on the reference position, both centers of the target and another off-slit point source were measured by the 2DGA.
This method is reliable without slit mask obstructions.
After moving the target onto the slit, we measured the center of the off-slit point source by 2DGA and derived the expected target position from the offset between the two sources.
Figure \ref{fig:TestProcess} shows the coordinates and definitions of position parameters.
The red dotted circles of (a) and (c) are target and off-slit point source. 
In (b), $x_{T}$ and $y_{T}$ are the center values of the target. 
$x_{G}$ and $y_{G}$ are the center values of the off-slit point source. 
These center positions were derived by 2DGA.
From the center values, we got a distance ${\Delta}X_{TG}$ and ${\Delta}Y_{TG}$ (the orange triangle and text).
After the target goes into the slit such as the blue and purple arrow (in case of A box, the red box) in (d), we inferred the expected center of the target ($x_{T}$=$x_{G}$+${\Delta}X_{TG}$, $y_{T}$=$y_{G}$+${\Delta}Y_{TG}$).
The center of the target with slit mask obstruction was measured by both 2DGA and CBA.
Note that the definitions of the centers from both 2DGA and CBA would be slightly different with saturated or non-symmetrical PSF.

\subsection{Center Finding Errors}

\begin{figure}
\epsscale{1.2}
\plotone{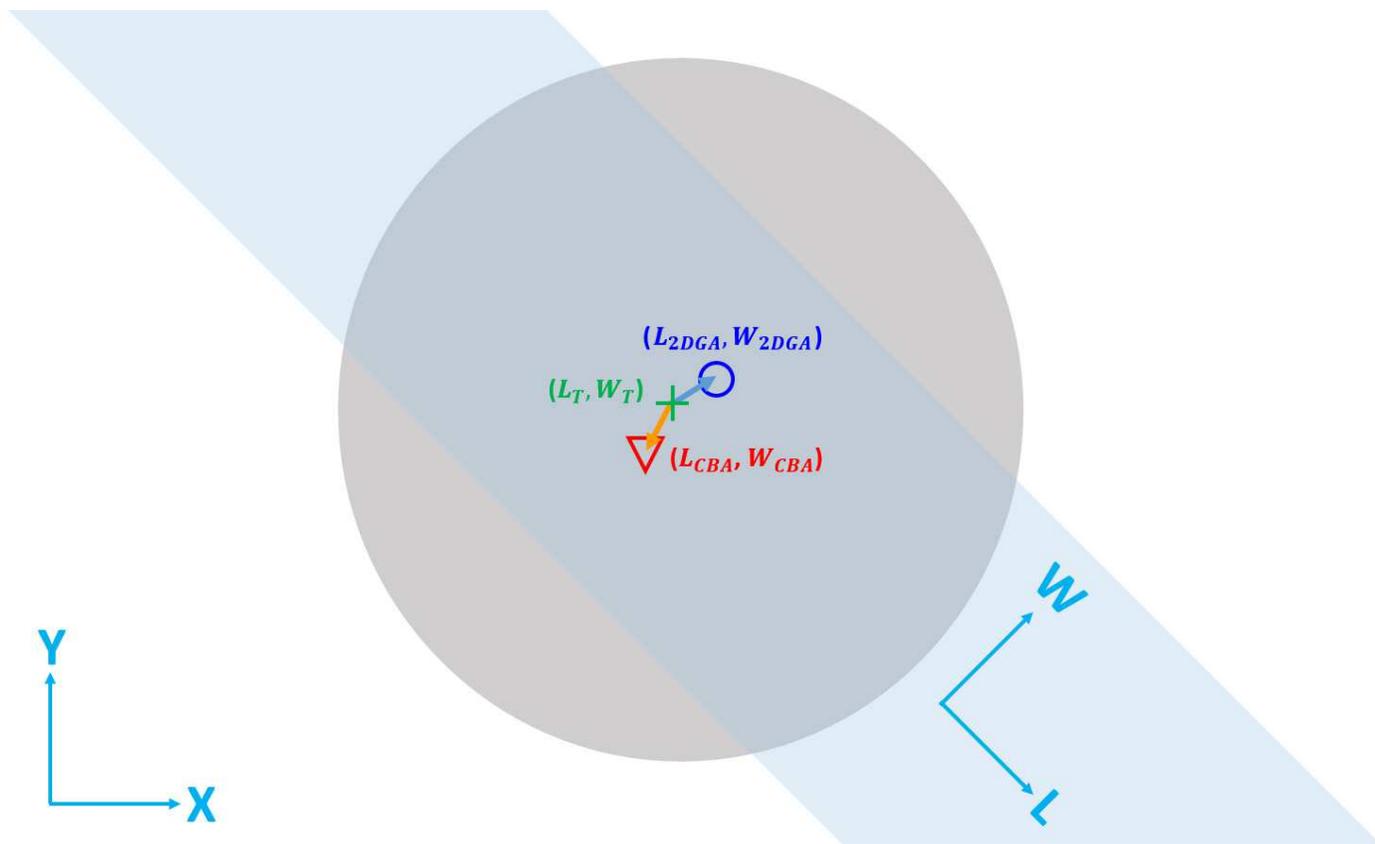}
\caption{Centers of the expected target position, and the measured target positions with slit mask obstruction (Figure \ref{fig:TestProcess}). 
This is the expanded target on the slit. 
The gray circle is the target and the sky color range is covered part by the slit.
We use both $X$ and $Y$ coordinate system and slit-length ($L$) and slit-width ($W$) coordinate system.
$L_{T}$ and $W_{T}$ (the green cross and text) are transformed to slit-length ($L$) and slit-width ($W$) coordinate system from the expected center ($x_{T}$, $y_{T}$) in Figure \ref{fig:TestProcess}. 
This figure shows the differences between $L_{2DGA}$ and $W_{2DGA}$ (the blue circle and text) by 2DGA and the expected center (the error: the cyan arrow) or between $L_{CBA}$ and $W_{CBA}$ (the red inverted triangle and text) by CBA and the expected center (the error: the orange arrow) on the SL$-$SW coordinate.
\label{fig:D_LW}}
\end{figure}

Since the center finding errors are mixed in the elements of $X$ and $Y$,
we have transformed the image data from the $X$ and $Y$ coordinate system in the array format to the slit-length ($L$) and slit-width ($W$) coordinate system (see Figure \ref{fig:D_LW}).
Since the pointing along the $W$ direction is more critical for minimizing slit-loss, only the slit-width direction component was considered \citep{iseki08}.
The measured target centers, with slit mask obstruction, from the 2DGA and the CBA, subtracted by the expected target center, i.e., (${\Delta}W_{2DGA}$ = $W_{2DGA} - W_{T}$) and (${\Delta}W_{CBA}$ = $W_{CBA} - W_{T}$) were the errors of the center finding algorithms (see Figure \ref{fig:D_LW}).

%%-------------------------------------------------------------------------------------------------------------

\section{Result and Discussion}

\subsection{Characterizing Error Patterns}

\begin{figure}
\epsscale{1.1}
\plotone{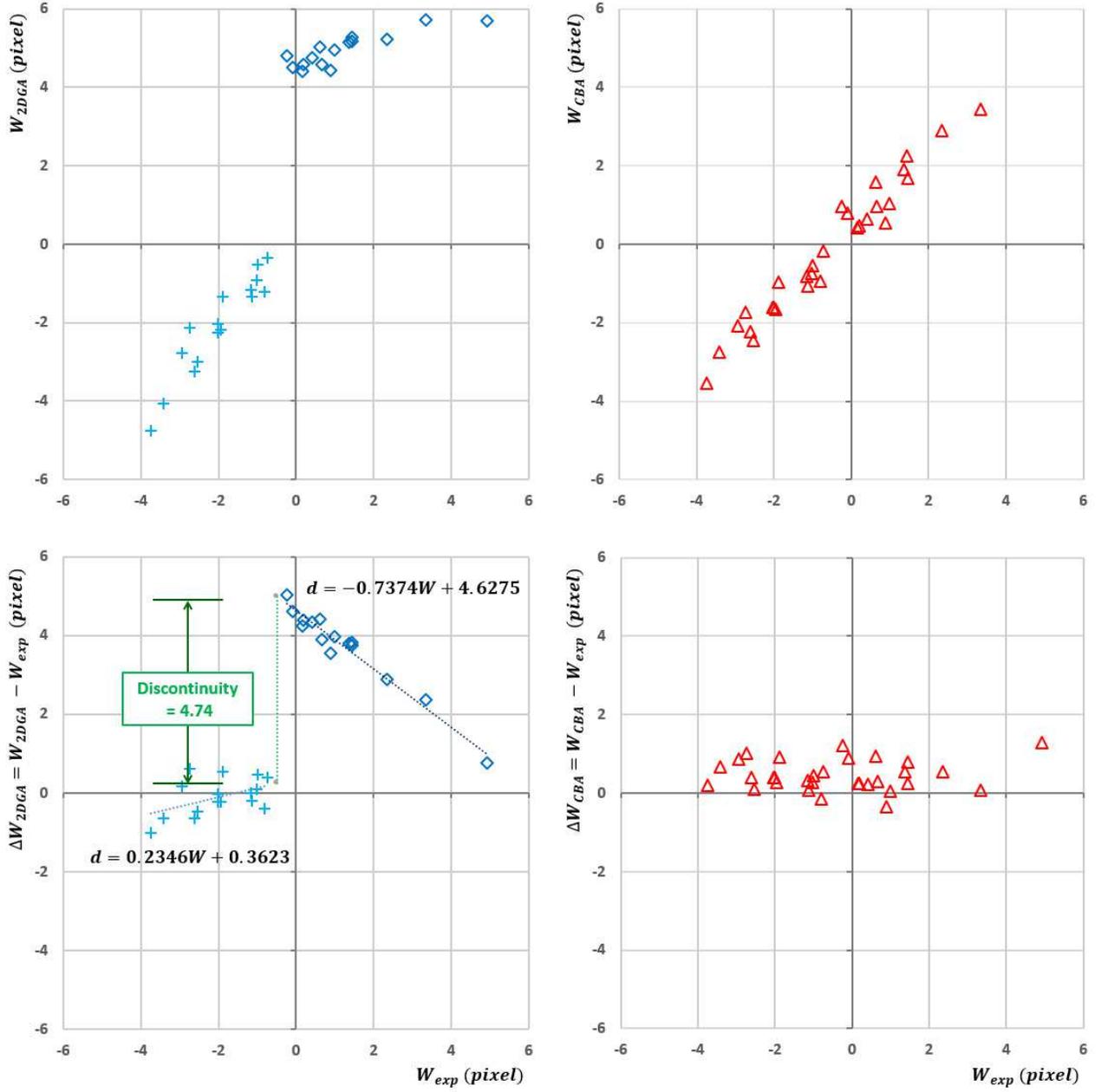}
\caption{%Slit distance VS. ${\Delta}_{Algorithm}$. 
Plots of the measured center positions ($W_{2DGA}$ and $W_{CBA}$ in the upper panels) and errors (${\Delta}W_{2DGA}$ and ${\Delta}W_{CBA}$ in the lower panels) versus the expected center positions ($W_{exp}$) along the slit-width direction.
The upper-left shows the center by the 2DGA along $W_{exp}$ (the blue diamond and the cyan cross), the upper-right is for the CBA (the red triangle).
The bottom-left plot shows the delta from the 2DGA to $W_{exp}$ (error: the cyan arrow in Figure \ref{fig:D_LW}), in which the discontinuity (the green box) is defined, and the bottom-right for the CBA (error: the orange arrow in Figure \ref{fig:D_LW}).
Data in the bottom-left plot can be fitted with linear functions $d(W) = 0.2346W + 0.3623$ (the cyan dotted line), $d(W) = -0.7374W + 0.3623$ (the blue dotted line), respectively. 
The discontinuity value {\it D} which is driven from these equations is 4.74 (the text inside green box).
This example is the same as target of Figure \ref{fig:TestProcess}.
\label{fig:TestProcess_Result}}
\end{figure}

Figure \ref{fig:TestProcess_Result} shows the measured center positions ($W_{2DGA}$ and $W_{CBA}$) and errors (${\Delta}W_{2DGA}$ and ${\Delta}W_{CBA}$) versus the expected center positions ($W_{T}$).
The distribution of the errors in the slit coordinate, as well as the root mean square (RMS) of the errors, can be compared to evaluate the performance of the center finding algorithms.

The errors from the 2DGA in Figure \ref{fig:TestProcess_Result} show a noticeable discontinuity pattern around the center.
This problem arises when the FWHM of the target PSF is smaller than the slit width.
Figure \ref{fig:slitmask} shows the slit-blocked PSF profiles, which the 2DGA may misidentify the edge of the slit aperture as the center of the PSF profile.
When the target is near this discontinuous position, the auto-guiding can be very unstable and begin jumping between the two slit edges.
To quantize this fatal feature, we modeled a simple linear discontinuous function, $d(W)$,

\begin{equation}
   d(W)=
    \begin{cases}
    a_{0} + a_{1}W, & \text{when $W \leq c$} \\
    b_{0} + b_{1}W, & \text{when $W > c$}
    \end{cases} 
\end{equation}

Where $a_{0}$, $a_{1}$, $b_{0}$, $b_{1}$, and $c$ values derive from the minimum $\chi$-squared value.
We define a discontinuity value, {\it D}, as follows:

\begin{equation}
    D = |a_{0} + a_{1}c - (b_{0} + b_{1}c)|
\end{equation}

The above discontinuity value is another way of showing the performance and stability of the center finding algorithms.
Figure \ref{fig:TestProcess_Result} shows that error (${\Delta}W_{CBA}$) distributions from the CBA do not have any discontinuity features.

\subsection{Errors in Various Conditions}

\begin{figure}
\epsscale{1.2}
\plotone{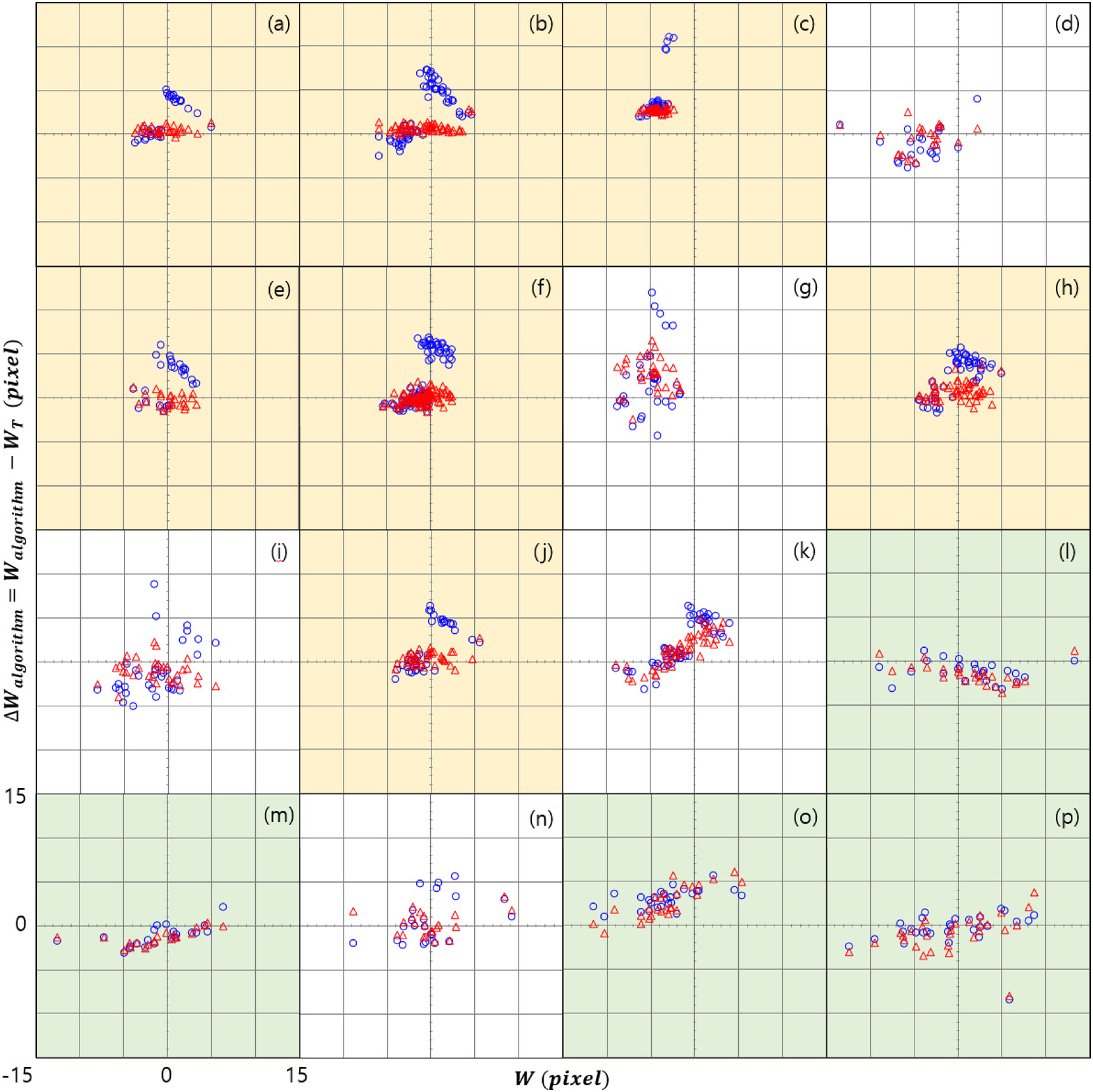}
\caption{Plots of errors (${\Delta}W_{2DGA}$ and ${\Delta}W_{CBA}$) versus $W$ from 16 samples. 
The blue circles are values from the 2DGA, the red triangles are measurements from the CBA.
%The sample target plots \textbf{are driven and overlapped using the way as the bottom plots in Figure \ref{fig:TestProcess_Result}}, which have reliable 2DGA results, are shown in green background.
Green backgrounds identify which tests have reliable 2DGA results and discontinuity features are shown with yellow backgrounds.
The target samples are listed in Table \ref{tab:result}. 
\label{fig:results}}
\end{figure}

In this section, we compare the performance of the 2DGA and the CBA in various observing conditions.
Figure \ref{fig:results} shows plots of errors for all our data from the 16 targets.
The error distribution patterns are different for each center finding algorithm, and depends on the target brightness and seeing conditions.
Table \ref{tab:result} lists RMS values from the 2DGA and the CBA, and the discontinuity values from the 2DGA.
The CBA does not display discontinuities.

\begin{figure}
\epsscale{1.1}
\plotone{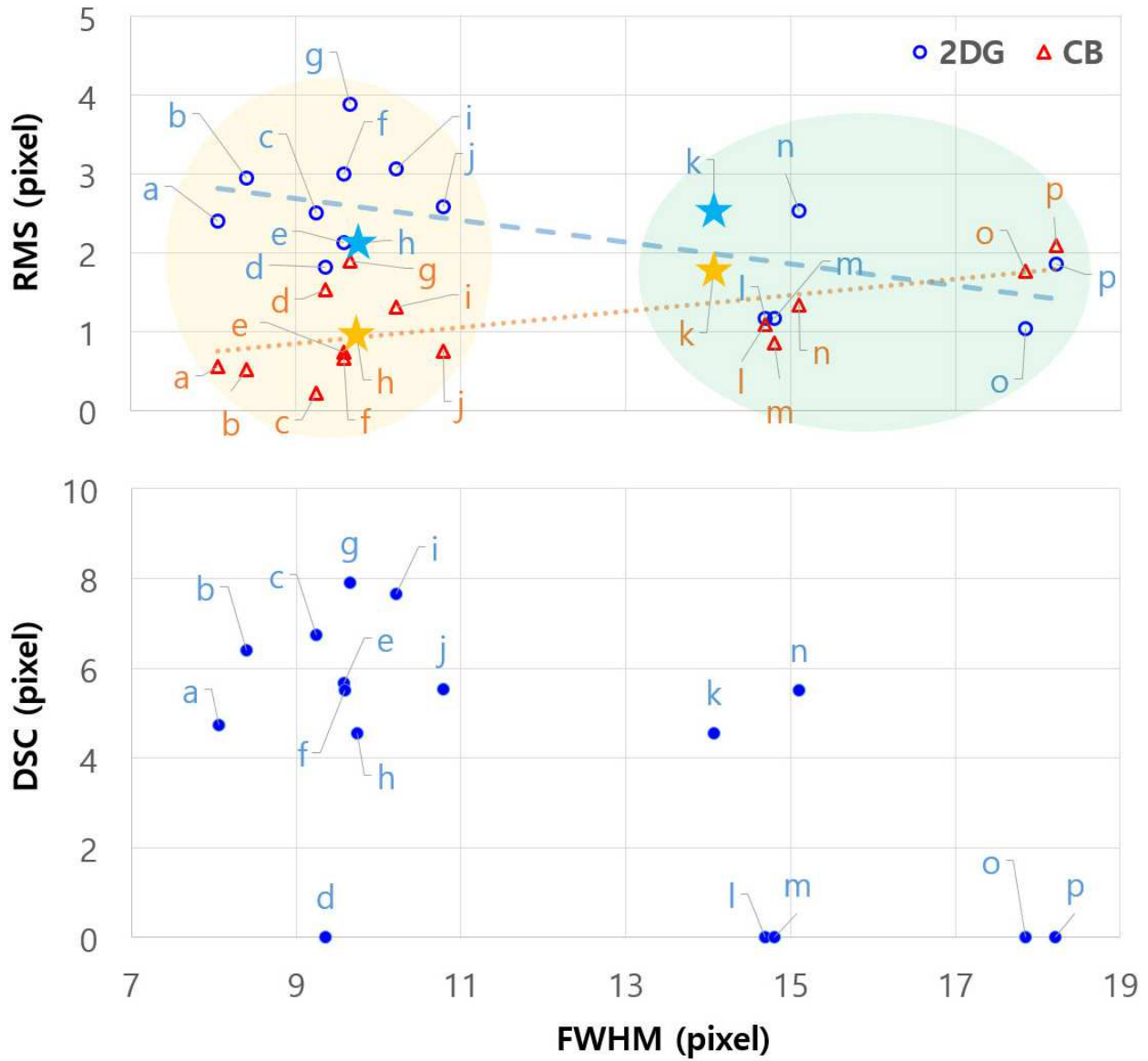}
\caption{Plots of RMS values and the discontinuity values versus FWHM of the PSF. 
From Figure \ref{fig:results}, we derived the relationship between the root mean square (RMS) of the errors and FWHM in the plot of the top. 
The plot of the bottom shows the discontinuity and FWHM (the blue filled circle).
The blue open circle and the red open triangle represent the results from the 2DGA and the CBA, respectively.
The green colored and yellow regions as same as Figure \ref{fig:results} shows reliable 2DGA results, samples with discontinuity features, respectively.
The blue dashed and orange dotted line is fitted based on 2DGA and CBA results for finding a tendency.
The same named labels (a to p) of circles and triangles are the same targets.
Table \ref{tab:result} lists the plotted values of the 16 target samples as the same labels.
Stars of same colors mean the same targets observed in two different nights (label h and k).
\label{fig:result_graph_FWHM}}
\end{figure}

We plotted the RMS values and the discontinuity values as a function of the FWHM of the PSF (see Figure \ref{fig:result_graph_FWHM}).
With smaller FWHM values, the RMS values from the 2DGA are larger, while those from the CBA are somewhat smaller.
To find the relationship between the error distribution patterns in Figure \ref{fig:results} and the RMS and discontinuity values in Figure \ref{fig:result_graph_FWHM}, we marked with the yellow background color on both figures.
When the FWHM is large (with the green background color), the RMS differences between the 2DGA and the CBA are not significant.
The CBA outperforms the 2DGA when the FWHM is small, which is also when the discontinuity feature in the 2DGA becomes 4-8 pixels ($\sim$0.5-1 slit width).
%However, the fatal discontinuity feature should be considered.
%Except for every large FWHM case, e.g., $\sim$18 pixels, most of the 2DGA results have small discontinuities.

\begin{figure}
\epsscale{1.1}
\plotone{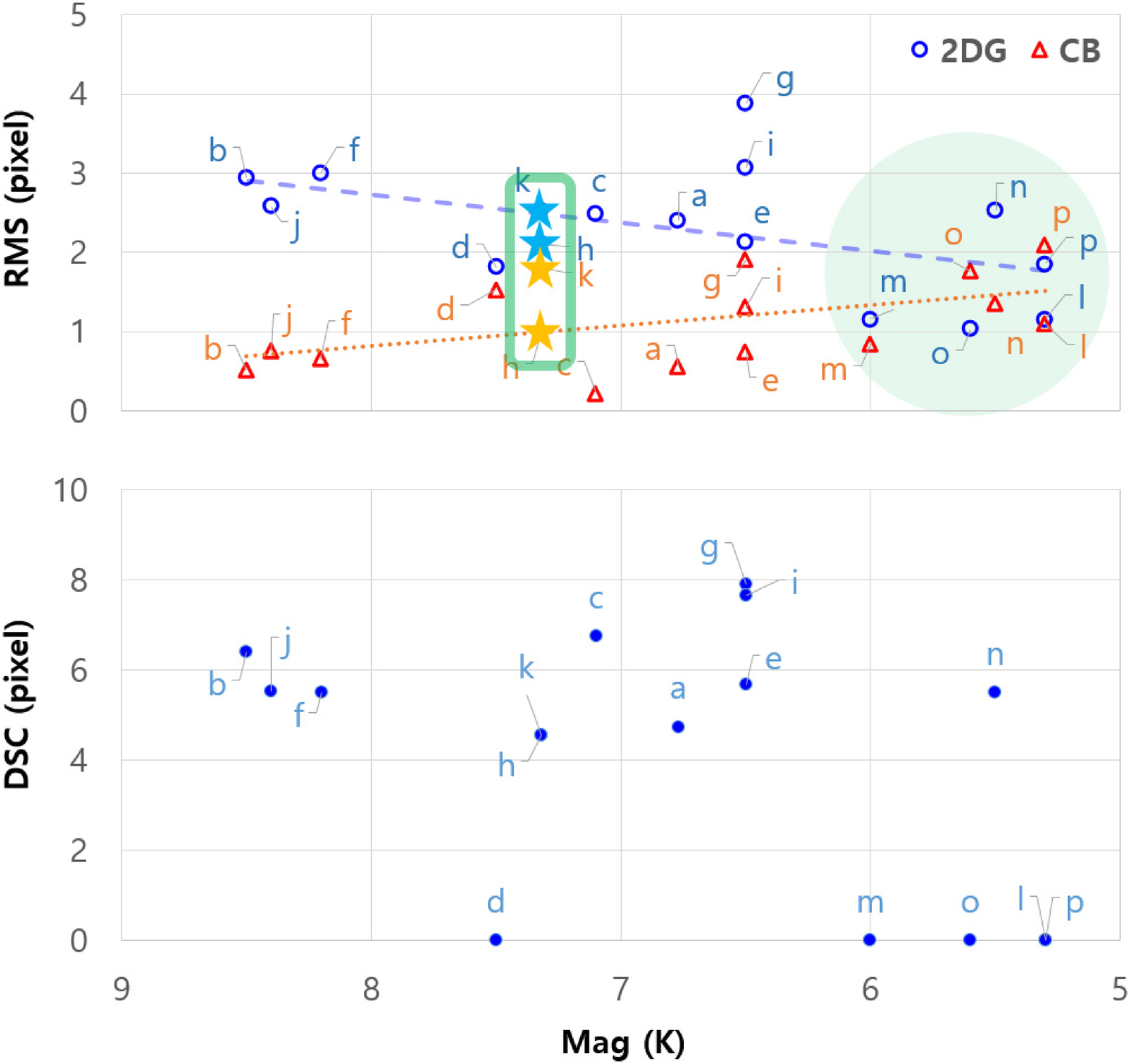}
\caption{Plots of RMS values and the discontinuity values versus the apparent $K$ magnitude.
As Figure \ref{fig:result_graph_FWHM}, we derived the relationship between the RMS of the errors and $K$ magnitude in the plot of the top.
The plot of the bottom shows the discontinuity and $K$ magnitude (the blue filled circle).
The blue open circle and the red open triangle represent the results from the 2DGA and the CBA, respectively.
The green colored regions shows reliable 2DGA results as same as Figure \ref{fig:results}.
The blue dashed and orange dotted line is fitted based on 2DGA and CBA results for finding a tendency.
The same named labels (a to p) of circles and triangles are the same targets.
Table \ref{tab:result} lists the plotted values of the 16 target samples as the same labels.
Stars of same colors mean the same targets in the other days (label h and k in the green box, which means same target).
\label{fig:result_graph_mag}}
\end{figure}

Figure \ref{fig:result_graph_mag} shows the performances as a function of the target magnitude.
Fainter targets have less overflow flux on the sides of the slit, which decreases the reliability of the 2DGA to fit the peak using the fainter wings of the PSF.
The RMS values from the 2DGA are slightly decreasing as the brightness of the target increases.
For very bright targets, e.g., $m_{K}$ < 6 mag, four out of five samples do not have discontinuity features
because saturation causes large FWHM values.
The CBA performs well, even at low flux, because the ratio of overflow flux is minimally impacted by a decrease in overall flux.
The RMS values from the CBA show different trends, performing slightly better for fainter targets, as we see in Figure \ref{fig:result_graph_FWHM}.

%%-------------------------------------------------------------------------------------------------------------

\section{Conclusions}

We have applied software engineering methods, i.e., the model-based design and the spiral development process, to IGRINS control software development.
To maximize the number of collecting photons in the spectrograph slit, we made CBA in addition to the typical 2DGA.
Applying both algorithms to commissioning observations, in various observing conditions at McDonald Observatory, we showed that they both perform well in poor seeing conditions.
Since very bright point sources are easily saturated at the peak, the size of the PSF becomes bigger even in good seeing conditions and this limits the effectiveness of the 2DGA.
When the FWHM is comparable to the slit width, the 2DGA algorithm shows a discontinuity near the slit center because most of the stellar flux passes through the spectrograph slit.
In typical observing conditions, the CBA finds the center of the slit-blocked image better than the 2DGA.

%%-------------------------------------------------------------------------------------------------------------

%\acknowledgments
\section*{Acknowledgement}
This work was supported by the National Research Foundation of Korea (NRF), grant No. 2017R1A3A3001362, funded by the Government of the Republic of Korea (MSIP).
Hye-In Lee and Huynh Anh N. Le were supported by the BK21 Plus program through the NRF funded by the Ministry of Education of Korea.
This work used the Immersion Grating Infrared Spectrometer (IGRINS) that was developed under the collaboration between the University of Texas at Austin and the Korea Astronomy and Space Science Institute (KASI) with the financial support of the US National Science Foundation under Grant AST-1229522 of the University of Texas at Austin, and of the Korean GMT Project of KASI. 
The IGRINS software Packages were developed based on the contract between KASI and Kyung Hee University. 
We thank the anonymous referees for their critical comments to improve this paper. 
We appreciate Dr. John Kuehne of the McDonald Observatory, Dr. Moo-Young Chun and Dr. Jae-Joon Lee of KASI, and Prof. Sungwon Lee and Mr. Bong-Yong Kwon of KHU for contributing to this research.
We also thank Prof. George Parks and Ms. Elaine S. Pak for proofreading this manuscript.
This paper includes data taken at the McDonald Observatory of the University of Texas at Austin.

%%-------------------------------------------------------------------------------------------------------------

%%-----------------------------------------------------------
%%tables

%%Table.1
\begin{deluxetable*}{crr}[b!]
\tablecaption{Balancing table of SS433 \label{tab:BT}}
\tablecolumns{15}
\tablewidth{0pt}
\tablehead{
\colhead{Index} & \colhead{Offset} & \colhead{$B$-ratio} \\
\colhead{} & \colhead{(pixel)} & \colhead{}
}
\startdata
1 & 31 & 0.1001 \\
2 & 32 & 0.1332 \\
3 & 33 & 0.1782 \\
4 & 34 & 0.2438 \\
5\tablenotemark{*} & 35 & 0.3451 \\
6 & 36 & 0.5054 \\
7 & 37 & 0.7527 \\
8\tablenotemark{*} & 38 & 1.1320 \\
9 & 39 & 1.7160 \\
10 & 40 & 2.5750 \\
11 & 41 & 3.8983 \\
12\tablenotemark{*} & 42 & 5.7586 \\
13 & 43 & 8.3880 \\
\enddata
%\tablenotetext{*}{}
\tablecomments{Sample data were taken on 2014 May 24. 
``Offset'' is the distance between a virtual slit and the center of the target.
``$B$-ratio'' is the balancing ratio of the slit-width direction between up and down from the virtual slit.
The virtual-slit-blocked PSF images and the PSF profiles of Index 5, 8, and 12 (marked with ``*'' symbols) are shown in Figure \ref{fig:slitmask}.}
\end{deluxetable*}

%%-----------------------------------------------------------
%%Table.2

% \usepackage{booktabs}
% \usepackage{multirow}
%%\begin{table}[]
\begin{deluxetable*}{llrrrrrc}[b!]
\tablecaption{Comparison of Results from 2DGA and CBA algorithms \label{tab:result}}
%%\begin{tabular}{@{}|l|l|r|r|r|r|r|c|@{}}
%%\toprule
\tablehead{
\multicolumn{1}{c}{\multirow{2}{*}{Index}} & \multicolumn{1}{c}{\multirow{2}{*}{Target}} & \multicolumn{1}{c}{\multirow{2}{*}{Mag (K)}} & \multicolumn{1}{c}{\multirow{2}{*}{FWHM (pixel)}} & \multicolumn{2}{c}{2DGA}                           & \multicolumn{1}{c}{CBA} & \multirow{2}{*}{Label} \\ \cmidrule(lr){5-7}
\multicolumn{1}{c}{}                       & \multicolumn{1}{c}{}                        & \multicolumn{1}{c}{}                         & \multicolumn{1}{c}{}                      & \multicolumn{1}{c}{Rms} & \multicolumn{1}{c}{DSC} & \multicolumn{1}{c}{Rms} &                         }
\startdata
20140525                                    & 2MASS J18331755                              & 6.8                                           & 8.06                                       & 2.40                     & 4.74                     & 0.55                     & a                      \\ 
20140526                                    & Serpens2                                     & 8.6                                           & 8.40                                       & 2.95                     & 6.40                     & 0.51                     & b                      \\ 
20140524                                    & Serpens15                                    & 7.0                                           & 9.24                                       & 2.50                     & 6.75                      & 0.22                     & c                      \\ 
20140711                                    & SR4 (V* V2058 Oph)                           & 7.5                                           & 9.36                                       & 1.81                     & 0                        & 1.53                     & d                      \\ 
20140713                                     & HD155379                                     & 6.5                                           & 9.58                                       & 2.13                     & 5.68                     & 0.74                     & e                      \\ 
20140524                                    & SS433                                        & 8.2                                           & 9.59                                       & 3.00                     & 5.51                     & 0.66                     & f                      \\ 
20140712                                    & HD155379                                     & 6.5                                           & 9.65                                       & 3.88                     & 7.90                     & 1.90                     & g                      \\ 
20140712                                    & GSS32                                        & 7.3                                           & 9.74                                       & 2.11                     & 4.56                     & 1.00                     & h                      \\ 
20140711                                    & HD155379                                     & 6.5                                           & 10.22                                      & 3.07                     & 7.65                     & 1.31                     & i                      \\ 
20140525                                    & V2247Oph                                     & 8.4                                           & 10.78                                      & 2.58                     & 5.54                     & 0.75                     & j                      \\ 
20140526                                    & GSS32                                        & 7.3                                           & 14.07                                      & 2.53                     & 4.55                     & 1.80                     & k                      \\ 
20140708                                     & HIP5131                                      & 5.3                                           & 14.69                                      & 1.16                     & 0                        & 1.09                     & l                      \\ 
20140525                                    & 25 Oph                                       & 5.5                                           & 15.10                                      & 2.53                     & 5.50                     & 1.34                     & n                      \\
20140709                                    & HIP95560                                     & 5.6                                           & 17.85                                      & 1.03                     & 0                        & 1.77                     & o                      \\ 
20140709                                    & HIP93580                                     & 5.3                                           & 18.21                                      & 1.86                     & 0                        & 2.09                     & p                      \\ \bottomrule
\enddata
%\end{tabular}
\tablecomments{``Index'' is a list of commissioning dates. 
``2DGA'' and ``CBA'' represent two-dimensional Gaussian fitting algorithm and center balancing algorithm, respectively.}
``FWHM'' is taken through 2DGA on the reference position.
``Label'' is a list of FWHM arranged in a large order which is used Figure \ref{fig:results}.
%\end{table}
\end{deluxetable*}

%%-----------------------------------------------------------
%%figure

\end{CJK}

\end{document}